# Recursive Random Number Generator Using Prime Reciprocals

Subhash Kak

## Introduction

Pseudorandom sequences have a variety of applications in computing, communications and in simulations. Amongst the methods used to generate such sequences are shift registers and linear congruential generators [1],[2].

I investigated the cryptographic properties of recurring sequences of digits obtained by prime reciprocals, $\frac{1}{p}$, also called d-sequences, many years ago in a series of papers [3],[4],[5]. As illustration, the prime reciprocal of 1/7 is $0.\dot{1}4285\dot{7}$, and the prime reciprocal sequence is considered to be 142857. The period of this sequence is 6. Since the maximum length of the period can be (*p-1*), this is an example of a maximum-length d-sequence. But a prime reciprocal sequence that is maximum length in one base need not be maximum length in another base. Thus the binary prime reciprocal sequence corresponding to 1/7 is 001, which has a period of 3. It follows from elementary mathematics that the period of a prime reciprocal must divide *(p-1)*.

The binary d-sequence is generated by means of the algorithm:

$$a(i) = 2^i \bmod p \bmod 2 \qquad (1)$$

where p is the prime number (for details, see [4],[5]). The maximum length (period *p-1*) sequences are generated when 2 is a primitive root of *p*. When the binary d-sequence is of maximum length, the bits in the second half of the period are the complements of those in the first half.

Examples of binary d-sequences;

for *p=13* → 000100111011

for *p=19* → 000110101111100101

Both of these are maximum length and one may check the asymmetry in 0s and 1s across half the period. It is easy to generate d-sequences, which makes them attractive for many engineering applications [6]-[10]. It should also be noted that any random sequence can be expressed as the rational number

$$\frac{a}{\prod_i p_i} \qquad (2)$$

where $a < \prod_i p_i$. For example, the shift register pseudo-random sequence 001101011110001 with period 15 which is generated by the feedback polynomial $x^4 + x^3 + 1$ in a linear feedback shift register can also be obtained by the rational number $\frac{6897}{32767} = \frac{3 \times 11 \times 209}{7 \times 31 \times 151}$. This means that we can consider d-sequences to be the most general representation of any random sequence.

Clearly, shift register pseudo-random sequences can be easily represented by d-sequences but the reverse is not necessarily true.

Note that the number of elements of the prime reciprocal sequence is much larger than the elements needed to represent the prime. The expansion in the number of elements if exponential and it is given by:

$$\frac{p-1}{\log_2 p} \qquad (3)$$

for a maximum-length, binary d-sequence. For other d-sequences also the expansion will be of the same order.

## Increasing the sequence period

It was shown in [4] that it is easy to find $i$ given $\log_2 p$ bits of $a(i)$. Therefore, d-sequences cannot be directly used in computationally secure random number generator (RNG) applications.

To make d-sequences more effective, one may wish to increase their period. This may be done in a variety of ways including splicing different d-sequences together or the use of sequences to different bases and then converting the digits to base 2.

By adding together two or more different d-sequences (obtained by using primes $p_1, p_2, \ldots$) mod 2, we are able to introduce non-linearity in the generation process, and the resulting sequence becomes a good candidate for use as random sequence.

For convenience, we will now consider only two terms in the sum. If the individual sequences are maximum length, then the period of the sum will be

$$lcm\ (p_1\text{-}1)\ (p_2\text{-}1) \qquad (3)$$

But for randomly chosen primes we do not know whether the component sequences were maximum length and, therefore, the actual period would be a divisor of *lcm ($p_1$-1) ($p_2$-1)*.

## A recursive generator

In the expansion (1) we implicitly consider the generator seed to be 2. We can now relax this requirement and take the seed to be equal to *S*, which is relatively prime to each $p_i$, and the order of *S* does not divide *($p_i$-1)* for all *i*. Consider the power-exponent RNG generates bits according to the algorithm:

$$a(0) = S \bmod p_1 \bmod 2 \oplus S \bmod p_2 \bmod 2$$
$$a(1) = S^2 \bmod p_1 \bmod 2 \oplus S^2 \bmod p_2 \bmod 2$$
$$a(2) = S^4 \bmod p_1 \bmod 2 \oplus S^4 \bmod p_2 \bmod 2 \quad (4)$$
……

where $\oplus$ means modulo 2 addition.

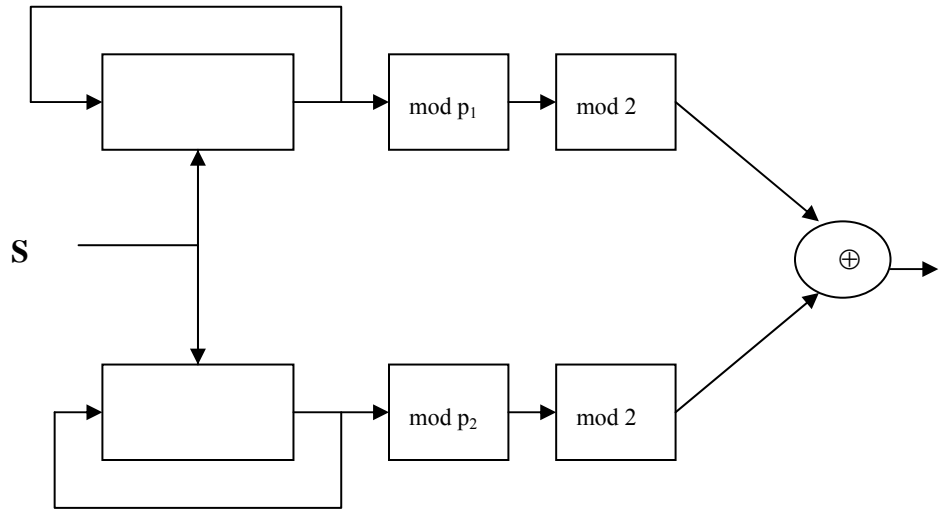

**Figure 1.** The schematic of the recursive random number generator

An even stronger RNG would be one where the two terms are added to different powers as shown below:

$$a(0) = S \bmod p_1 \bmod 2 \oplus S^k \bmod p_2 \bmod 2$$
$$a(1) = S^2 \bmod p_1 \bmod 2 \oplus S^{2k} \bmod p_2 \bmod 2$$
$$a(2) = S^4 \bmod p_1 \bmod 2 \oplus S^{4k} \bmod p_2 \bmod 2 \quad (5)$$
……

Since the seed *S* would be randomly chosen, the period of the sequence will be less than *lcm ($p_1$-1) ($p_2$-1)* if it is not a primitive root simultaneously of $p_1$ and $p_2$.

One may replace $p_1$ and $p_2$ by $n_1$ and $n_2$ that are product of primes. For better security, the two primes should each be congruent to 3 mod 4.

Exponents other than 2 of generators (4) and (5) may also be used.

## Discussion

There are many applications where the flexibility in the choice of literally any period for the pseudorandom sequence is an important consideration. For such applications the use of prime reciprocal sequences can be more convenient than the use of shift register sequences.

In problems dealing with the design of optimal sequences for specific communications applications, the use of prime reciprocals can likewise facilitates the process of finding and designing these sequences.